\documentclass{optica-article}

\journal{opticajournal} 

\articletype{Research Article}

\usepackage{lineno}
\usepackage{siunitx}
\usepackage{upgreek}

\begin{document}

\title{Portable laser-cooled ytterbium beam clock based on an ultra-narrow optical transition}

\author{
R. F. Offer,\authormark{1} 
E. Klantsataya,\authormark{1} 
A. P. Hilton,\authormark{1} 
A. Strathearn,\authormark{1} 
N. Bourbeau H\'{e}bert,\authormark{1}
C. J. Billington,\authormark{1} 
S. Watzdorf,\authormark{1} 
S. K. Scholten,\authormark{1} 
B. White,\authormark{1} 
M. Nelligan,\authormark{1} 
T. M. Stace,\authormark{2}
and A. N. Luiten\authormark{1,*}}

\address{\authormark{1}Institute for Photonics and Advanced Sensing (IPAS) and School of Physics, Chemistry and Earth Sciences, University of Adelaide, Adelaide, SA 5005, Australia\\
\authormark{2}School of Mathematics and Physics, University of Queensland, Brisbane, QLD 4072, Australia}

\email{\authormark{*}email: andre.luiten@adelaide.edu.au} 


\begin{abstract*} 
The highest performance atomic clocks are based on interrogation of ultra-narrow optical transitions.  
There is now significant interest in developing these systems as a source of GNSS-independent time in deployed, dynamic environments. 
We report on the development and field trial of a portable optical atomic clock interrogating the $10$\,mHz wide $^1$S$_0\rightarrow ^3$P$_0$ transition in ytterbium-171. 
To enable measurement of this ultra-narrow transition in a deployed setting we combine an atom-vapor based pre-stabilization reference with all-digital control and continuous clock spectroscopy of a transversely-cooled thermal atomic beam. 
Characterization of the short-term frequency stability within the lab demonstrates a modified Allan deviation of $2\times 10^{-14}/\sqrt{\tau}$ for integration times up to $100\,$s, reaching a best performance of $1.9\times 10^{-15}$ at $200\,$s.
The clock demonstrated the same performance after transport and install aboard a ship for field trial, and operated uninterrupted for multiple days whilst at sea.  
These results show a pathway towards truly portable optical frequency references based on the interrogation of ultra-narrow transitions.

\end{abstract*}

\section{Introduction} \label{Sec:Intro}

Optical atomic clocks have revolutionized the field of frequency metrology \cite{Ludlow2015}.
Interrogation of ultra-narrow optical transitions ($<$Hz linewidth) allows for orders of magnitude improvement in both accuracy and frequency stability compared to devices based on microwave transitions \cite{5071A,MHM2020}. 
There is now significant interest in developing optical clocks for operation outside of specialized lab environments in more deployed, dynamic settings \cite{Marlow2021, Mehlstaubler2018, White2024}. 
This is particularly motivated by a need for high-performance GNSS-independent timing sources for integration within both civilian and defence infrastructure \cite{npl2024}.  

In this work we present a new design of portable optical clock based on interrogation of the 10-mHz-wide ytterbium clock transition in a laser-cooled atomic beam.
This system aims to provide better frequency stability than the most precise commercial frequency standard, the hydrogen maser \cite{MHM2020}, whilst also operating in harsh environments.  
In operating this clock at sea, we have demonstrated robust interrogation of an ultra-narrow optical transition on a moving platform for the first time.

The best lab-based optical clocks interrogate trapped and cooled atoms or ions with Hz-level linewidth lasers, targeting ultra-narrow transitions with inherently low sensitivity to magnetic field and black body radiation \cite{Bothwell2022, Aeppli2024}.  
This requires a complicated vacuum system with multiple cooling and trapping lasers \cite{Katori2011, Margolis2009} and a high-finesse, environmentally-isolated optical cavity for pre-stabilization of the clock laser \cite{Robinson2019,Jiang2011}.  
These technical challenges have limited portable optical clocks based on cold samples to large, multiple rack systems that require significant set-up time \cite{PTB2018, RIKEN2020, Opticlock2021, Wuhan2020, LENS2014, Bothwell2025}.  
To our knowledge, all such demonstrations have required the clock to be stationary during operation.

Recently a new generation of vapor-based optical atomic clocks have been developed and demonstrated in field trials.
These systems, based on ytterbium \cite{Hilton2024}, rubidium \cite{Martin2018,Perrella2019}, and iodine vapors \cite{JOKARUS2019, Roslund2024}, are of a significantly simpler design, utilizing broader optical transitions ($\sim\,$MHz) within a thermal atomic vapor.
This removes the need for laser-cooling or pre-stabilization of the clock laser and allows for impressively robust operation.  However, this has come at the expense of reduced performance and an increased sensitivity to environmental effects. 

In contrast, we have developed a clock design that allows for interrogation of an ultra-narrow transition in deployed settings. 
We utilize Ramsey--Bord\'e spectroscopy \cite{Borde1984} of a collimated thermal atomic beam to produce a Doppler-free interferometric measurement of the $^1$S$_0 \rightarrow$ $^3$P$_0$ transition in neutral $^{171}$Yb. 
Previous lab-based optical beam clocks have been based on Mg \cite{Sengstock1994}, Ca \cite{Olson2019, McFerran2010, Riehle1992, Tobias2025} and Sr \cite{Fartmann2024}, with clock transition linewidths of 31\,Hz, 400\,Hz and 7\,kHz, respectively.
These clocks have demonstrated fractional frequency stabilities of better than $3\times 10^{-16}$ for timescales up to 3000 seconds \cite{Olson2019}. 
Our system is the first optical beam clock to be based on a clock transition with $<1\,$Hz linewidth, and the first system to allow interrogation of such a transition within a dynamic environment.  
The use of an atomic beam is not only significantly simpler than a fully cooled system, but also allows continuous read out of the clock signal.
This results in a locking bandwidth of $\sim100\,$Hz and makes the clock inherently more compatible with noisy environments. 
Furthermore, we tailor the design to allow operation without a pre-stabilization cavity, instead using spectroscopy of the relatively narrow $^1$S$_0 \rightarrow$ $^3$P$_1$ transition in a separate thermal Yb vapor cell as a robust frequency reference.

In the following we report on the operation and performance of the Yb beam clock within the lab and during an at-sea field trial carried out off the coast of Australia.  
This trial took place on board a vessel provided by the Royal Australian Navy and was organized by the Australian Government Department of Defence. 
In the lab we measure a clock fractional frequency instability of $2\times 10^{-14}/\sqrt{\tau}$ to 100 seconds of integration, with characterization at longer timescales limited by the frequency references available. 
The clock demonstrated the same level of performance after transport and install for the field trial, and operated uninterrupted for multiple days whilst at sea.
Using inertial data acquired during the trial we have measured the clock frequency sensitivity to rotation and acceleration, which is in close agreement with a theoretical model of the system. 

\section{Results} \label{Sec:Results}

\subsection{Spectroscopy of a 10\,mHz transition on a thermal atomic beam}

\begin{figure}[t] 
\centering
\includegraphics[width=0.6\linewidth]{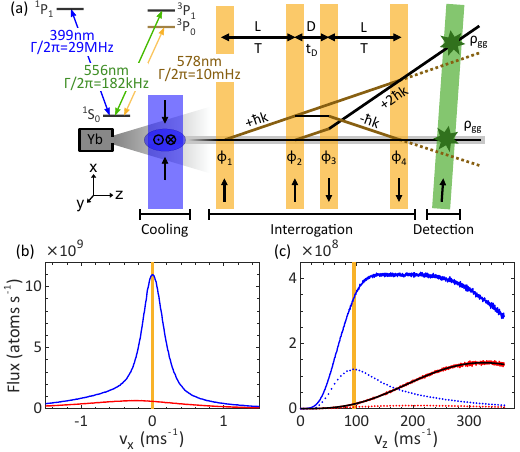}
\caption{\label{Fig:Concept} (a) Schematic of the clock physics package.  Atomic beam passes from left to right through cooling, interrogation, and detection lasers, which address the 399\,nm, 578\,nm, and 556\,nm transitions in neutral $^{171}$Yb, respectively. Experimental measurement of the atomic flux distribution as a function of (b) transverse and (c) longitudinal velocity as measured in the detection stage, without (red) and with (blue) cooling.  The dotted lines show the longitudinal velocity distribution which is excited during RB interrogation (see Supplementary Material Sec.\,1 for details), and the yellow shading indicates the range of velocities that contribute to the clock measurement.  The black curve in (c) shows a fit to a Maxwell--Boltzmann flux distribution with a temperature of $454^{\circ}$C.  }
\end{figure}

Clock interrogation within the beam clock is performed via Ramsey--Bord\'e (RB) spectroscopy.  
As shown in Fig.\,\ref{Fig:Concept} (a), the atomic beam passes through two pairs of counter-propagating clock laser beams, with the atoms ideally experiencing a $\pi/2$ pulse at each interaction.
This splits the atomic trajectories, forming two closed matter-wave interferometers \cite{Borde1989}, and produces a Doppler--free interferometric comparison between the clock laser frequency and the atomic transition frequency.  
The resulting interference fringes in the ground state atomic population are of the form \cite{Borde1984}
\begin{equation}
\rho_{gg} \propto \text{cos}(4\pi T(\Delta \pm \delta)+\phi)
\end{equation}
where T is the time of flight within each Ramsey zone and $\delta$ is the recoil shift.  
The laser phases $\phi_i$ at each interaction zone determine the net phase $\phi$ of the fringes, with $\phi=\phi_2-\phi_1+\phi_4-\phi_3$.
The fundamental performance of a clock based on this technique is optimized by maximizing the atom flux contributing to the RB fringe whilst minimizing noise-contributing background atomic flux. 

Implementation of this technique for an ultra-narrow transition on a thermal atomic beam requires careful consideration.
The principal challenge is the extremely low coupling strength of the transition, and correspondingly long interaction time or high laser power required to produce sufficient excitation of the clock state.

The accumulated pulse area as the atoms pass through each laser is given by \cite{Strathearn2023, Foot2004}
\begin{equation}
\Phi = \frac{\Gamma}{v_z} \sqrt{\frac{P_0}{I_{\text{sat}}}}\text{exp}[-\frac{(\Delta-k v_x)^2 \tau^2}{4}]
\end{equation}
where $v_z$ is the atomic velocity along the atomic beam, $v_x$ is the atomic transverse velocity, $\tau$ is the crossing time of the atoms through the laser, $\Delta$, $P_0$ and $k$ are the laser detuning, power and wavenumber, and $I_{\text{sat}}$ and $\Gamma$ are the transition saturation intensity and linewidth. 
For reasonable laser powers ($<500$\,mW) this limits the maximum longitudinal atomic velocity that can receive a $\pi/2$ pulse on the $^1$S$_0\rightarrow$ $^3$P$_0$ transition to $\sim110$\,m$\text{s}^{-1}$.
Choosing a beam diameter of $1\,$mm sets the Fourier-broadened linewidth of the transition and limits the maximum transverse atomic velocity to $0.015$\,m$\text{s}^{-1}$.
These velocity ranges are very poorly matched to the typical flux distribution of an Yb beam, which has a most probable longitudinal velocity of around $320$\,m$\text{s}^{-1}$ (oven temperature of $450^{\circ}$C), and FWHM transverse velocity spread of around $30$\,m$\text{s}^{-1}$ (oven collimation of 100\,mrad).
In combination, this means that a fraction of less than $10^{-6}$ of $^{171}$Yb atoms produced by the oven are able to interact with the clock lasers, leading to a very poor signal-to-noise ratio (SNR).

We solve this problem by combining laser cooling and velocity-selective detection to increase the atomic flux within these velocity ranges and decrease measurement noise due to background ground state atoms.  
The atomic flux distribution as measured in the detection stage as a function of transverse atomic velocity is shown in Fig.\,\ref{Fig:Concept} (b) (see Supplementary Material Sec.\,1 for details). 
Apertures along the path of the atomic beam limit the measured FWHM without cooling (red) to $1.8$\,m$\text{s}^{-1}$.  
With 2D transverse cooling (blue), which is produced by two pairs of retro-reflected perpendicularly linearly polarized beams addressing the strong $^1$S$_0\rightarrow$ $^1$P$_1$ transition, the flux within the required velocity range for clock interrogation (yellow shading) is increased by a factor of 18. 

\begin{figure*} [b]
\includegraphics[width=\linewidth]{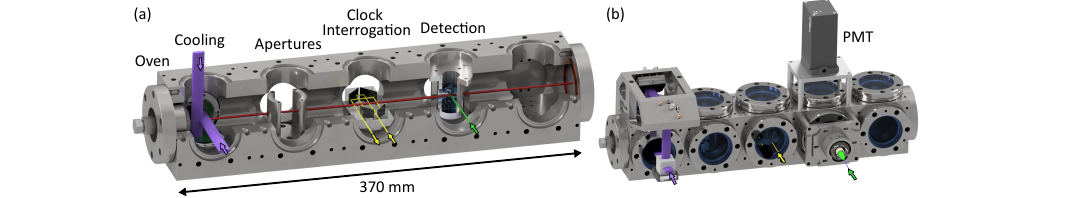}
\caption{\label{Fig:Chamber} Portable vacuum system for the Yb beam clock (a) showing the in-vacuum optical components of the cooling, interrogation and detection stages, and (b) showing the externally mounted PMT and cooling optics.  An off-the-shelf vacuum pump (not shown) maintains a base pressure of $1.5\times 10^{-9}$\,Torr.}
\end{figure*}

The cooling stage also changes the longitudinal velocity distribution of the atoms, significantly weighting the distribution to slower velocities, as shown in Fig.\,\ref{Fig:Concept} (c).
The dotted lines indicate the expected excited atom distribution after the full clock interrogation stage, both with and without cooling.
With cooling there is a significant increase in excited flux but measurement of the full atomic population would be limited by atom shot noise from the remaining population. 
To reduce this effect, we make the detection stage sensitive to longitudinal velocity to allow selective measurement of only the slower atoms. 
We carry out a ground state fluorescence measurement on the relatively narrow $^1$S$_0\rightarrow$ $^3$P$_1$ transition, and angle the excitation beam at a small angle $\theta$ from the perpendicular to introduce a Doppler shift of
\begin{equation}
\Delta_{v_z}=2\pi v_z \text{sin}(\theta)/\lambda.
\end{equation}
We then detune the detection laser such that fluorescence is driven only from a narrow velocity class ($7$\,m$\text{s}^{-1}$ FWHM at $\theta=2^{\circ}$ for the 182\,kHz linewidth of the detection transition, yellow shading Fig.\,\ref{Fig:Concept} (c)).
The net result of the cooling stage and velocity-selective readout is to increase the atom flux contributing to the fringe from $\sim 8\times 10^5$\,s$^{-1}$ to $1 \times 10^7$\,s$^{-1}$, whilst halving the flux of background atoms. 
This results in an expected increase in the SNR of the clock measurement by a factor of 19 (Supplementary Material Eq. 2). 

The vacuum system constructed to realize this scheme in a portable package is shown in Fig.\,\ref{Fig:Chamber} (a) and (b).   
Mechanical stability is maximized by machining the chamber from a single piece of stainless steel, and where possible optical components are mounted in vacuum.
The atomic beam is produced by an externally heated oven \cite{White2025} which operates at $450^{\circ}$C with a power draw of around 30\,W.
This is immediately followed by the 2D transverse cooling stage which is formed by in-vacuum retro-reflection optics. 
The atomic beam then passes through two 2\,mm diameter apertures before entering the clock interrogation stage where an interferometrically-aligned optical assembly produces the four RB beams from a single input beam.  
The mechanical robustness of this component is key to the clock performance (due to the dependence of the RB fringe phase on the net laser phase $\phi$) and it is constructed to meet stringent alignment requirements \cite{Klantsataya2025}.
Finally, the atoms propagate to the detection stage, where an in-vacuum lens and mirror assembly provides high-numerical aperture collection of ground state fluorescence onto the photo-multiplier tube (PMT) mounted externally at the top of the chamber.  
This assembly is able to capture an estimated 20\% of fluorescence produced in the center of the beam.

\begin{figure*}[t]
\centering
\includegraphics[width=\linewidth]{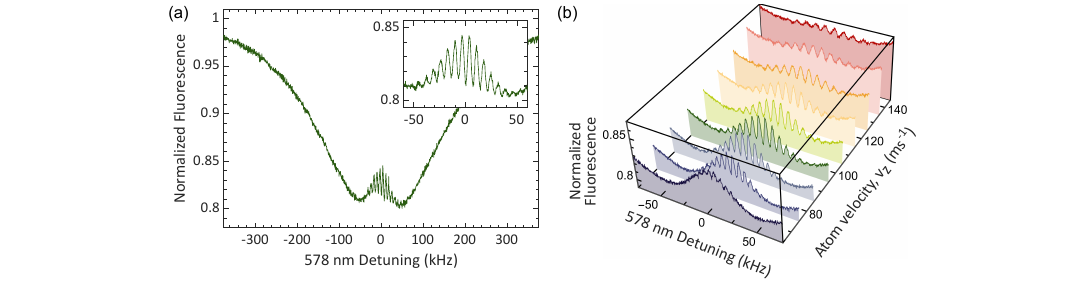}
\caption{\label{Fig:Spectroscopy} (a) RB spectroscopy signal for a longitudinal atomic velocity of 95\,m$\text{s}^{-1}$ and (b) RB spectroscopy over a reduced scan range for different longitudinal velocities. }
\end{figure*}

Typical spectroscopy of the clock transition obtained using this system is shown in Fig.\,\ref{Fig:Spectroscopy} (a), where we select a detection velocity class centered on $95\,$m$\text{s}^{-1}$.  
The Doppler-broadened background has a FWHM of 210 kHz, corresponding to a transverse velocity spread of 0.12\,m$\text{s}^{-1}$ FWHM, well below the Doppler limit of 0.3\,m$\text{s}^{-1}$.  
Due to the relatively narrow range of longitudinal velocities selected in the detection stage, multiple periods of RB fringe are observed.
As a result, the Ramsey length, L, is chosen so that the fringe period for the target velocity class is an integer multiple of the recoil shift, ensuring constructive interference of fringes from the two recoil components (Eq.\,1). 
Our chosen Ramsey length of $7\,$mm produces a measured fringe period of 7.0\,kHz $= 2 \delta$, where $\delta=3.5\,\mathrm{kHz}$. 

By choosing different detection detunings we are able to record RB spectroscopy for different longitudinal velocity classes, as shown in Fig.\,\ref{Fig:Spectroscopy} (b).
As expected from Eq.\,2, the fringe contrast decreases for increasing velocity due to the reduced pulse area. 
The reduction in fringe contrast for slower atoms is due to destructive interference of recoil components associated with the reducing fringe period.  

\subsection{Robust pre-stabilization} \label{Sec:Prestabilisation}

Robust pre-stabilization references for portable optical clocks is a growing area of research, with recent advances including commercially available cubic cavities \cite{Hill2021}, mm-scale vacuum gap references \cite{Kelleher2023, McLemore2024} and solid-state resonators \cite{Loh2020, Stern2020}.  
We choose an alternative approach, instead making use of the atoms themselves for pre-stabilization, which provide an absolute reference with inherently reduced inertial and environmental sensitivities.  

\begin{figure*}[t] 
\centering
\includegraphics[width=\linewidth]{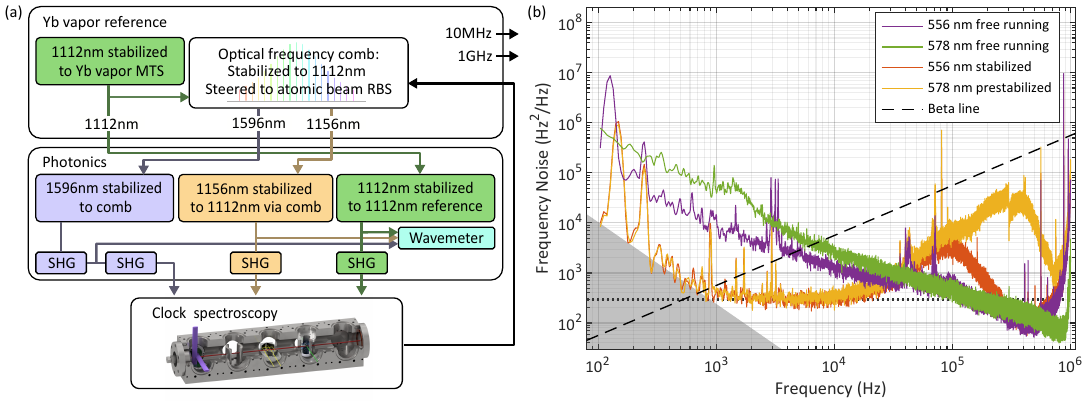}
\caption{\label{Fig:Photonics}(a) Schematic of the photonics subsystem of the clock and (b) frequency noise power spectral density of the 1112\,nm laser within the vapor reference and the 1156\,nm clock laser, scaled to their frequency doubled outputs.  Shaded gray region indicates noise floor of the interferometer.  Sharp tones from 100\,Hz to 1\,kHz in the stabilized data are produced by acoustic noise within the interferometer.} 
\end{figure*}

Fig.\,\ref{Fig:Photonics} (a) shows the layout of the photonics system integrated within the clock, comprising the vapor pre-stabilization reference and three fiber lasers that are frequency-doubled (or quadrupled) to produce the necessary cooling, interrogation and detection light. 
The vapor reference is similar to the Yb vapor clock demonstrated in \cite{Hilton2024}, and contains an 1112\,nm fiber laser which is frequency-doubled and stabilized to the $^1$S$_0 \rightarrow ^3$P$_1$ transition in $^{174}$Yb via modulation transfer spectroscopy of a thermal vapor.
It also includes a compact fiber-based frequency comb \cite{Sinclair2015}, which is self-referenced via f-2f stabilization and offset locked to the stable 1112\,nm laser via a phase-locked loop (PLL).

We characterize the frequency noise of the vapor reference and clock laser using a 200\,m path imbalanced fiber Mach--Zehnder interferometer \cite{Chen1989, Kefelian2009}.  
Both the 1112\,nm laser within the vapor reference and 1156\,nm clock laser are frequency-doubled before addressing their relevant atomic transition.  
We carry out the frequency noise characterization in the IR (with 1112\,nm and 1156\,nm outputs) and multiply the measured laser frequency time series by a factor of two to convert to frequency fluctuations at 556\,nm and 578\,nm, respectively.  
The results of this are shown in Fig.\,\ref{Fig:Photonics} (b).
Once stabilized, the 556\,nm light within the vapor reference achieves a white frequency noise floor of $300\text{Hz}^2/\text{Hz}$ (dotted line), with all control loop resonance peaks well below the beta line (dashed line) \cite{Di_Domenico2010}.  
This stability is transferred to the clock laser via a transfer lock using 1112\,nm and 1156\,nm outputs of the integrated frequency comb \cite{Hebert2020, Hebert2021}.
The resulting Lorentzian linewidth of the frequency-doubled clock laser due to white frequency noise is $\nu_{1/2}=940$\,Hz \cite{Di_Domenico2010}.  
For a Ramsey time of $T=74$\,$\upmu$s (corresponding to a $7$\,kHz fringe period) we expect this to decrease the observed fringe amplitude by a factor of $\text{exp}(-2\pi \nu_{1/2} T)=0.64$ \cite{Sengstock1994}.  

The vapor reference serves not only to suppress clock laser frequency noise, but also as a reference for the cooling and detection lasers.  
The detection laser (frequency-doubled 1112\,nm) is stabilized directly to the stable 1112\,nm laser within the vapor reference, with an offset frequency of $\sim 2$\,GHz due to the different isotopes used within the vapor and beam, whilst the cooling laser (frequency quadrupled 1596\,nm) is stabilized to an appropriate mode within the frequency comb. 

To stabilize the clock laser (and frequency comb) to the RB spectroscopy signal, we add slow steering to the offset between the frequency comb and the 1112\,nm vapor reference laser.
This produces a stable optical output at 1156\,nm, and stable 10\,MHz and 1\,GHz outputs are synthesized from the pulse train of the comb.
The steering signal is produced by modulating the clock laser frequency (modulation depth 1.5\,kHz, rate 1.5\,kHz) and demodulating the resulting RB spectroscopy signal with a bandwidth of 75\,Hz.  

All laser frequency locks are implemented digitally on five field programmable gate array (FPGA) signal processing units.
These individual systems are managed by an oversight PC embedded within the clock.

\subsection{Lab performance} \label{Sec:Lab}

\begin{figure*}[t] 
\centering
\includegraphics[width=\linewidth]{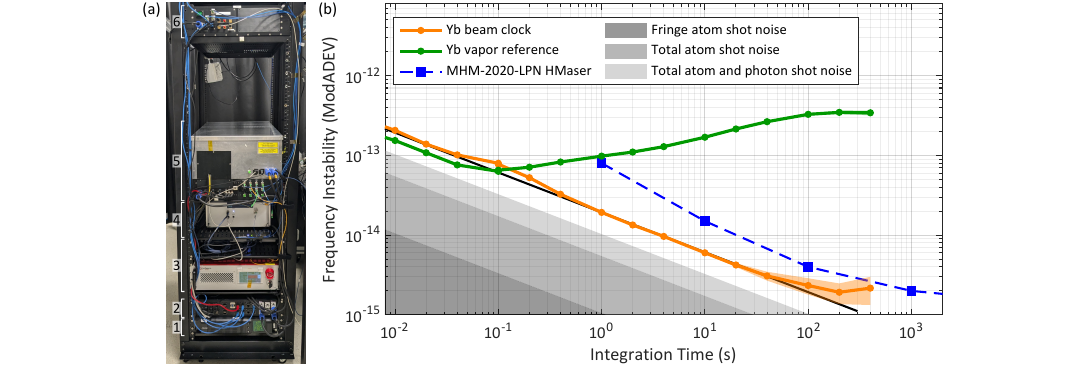}
\caption{\label{Fig:Lab}(a) The portable Yb beam clock, comprising 1. Uninterruptible power supply, 2. Power and oversight system, 3. Commercial cooling laser, 4. Commercial fiber amplifier for interrogation laser, 5. Physics package, 6. vapor reference. (b) Frequency instability ModADEV of the Yb beam clock (orange) and vapor reference (green) in lab-based tests, with specified frequency stability of the Microchip MHM-2020 hydrogen maser for comparison.  Black line shows $2\times 10^{-14}/\sqrt{\tau}$.  Shaded regions indicate stability limits due to predicted atom and photon shot noise from measured atomic fluxes, fluorescence and background scattering.}
\end{figure*}

As a complete device, the Yb beam clock is rack-mounted, totaling 25\,U in an 800\,mm deep rack (0.5\,$\text{m}^3$).  
It is powered by a single standard wall socket with an operational power draw of 770\,W, and approximate total weight of 150\,kg.
The majority of the clock is integrated within a single 7\,U package labeled the physics package in Fig.\,\ref{Fig:Lab} (a). 
This package is formed of two levels.  
The lower level houses the fiber photonics components and FPGA-based control systems for laser frequency and power stabilization. 
The upper level includes the vacuum chamber and all free-space optics, as well as the laser heads for the cooling and interrogation lasers.  
The remainder of the clock includes: the vapor reference, commercial laser systems for the cooling and interrogation lasers and a power and oversight package containing power supplies and the oversight PC.
The largest contributions to the total 770W power draw are the cooling and interrogation commercial laser systems, which consume 140\,W and 100\,W, respectively. 

The lab-based frequency stability has been characterized by making frequency comparisons to other portable optical clocks under development at the University of Adelaide (see Supplementary Material Sec.\,2 for measurement details).  
To assess the bandwidth and gain of the steering to the RB spectroscopy, we measure the stability of both the 1112\,nm output of the vapor reference (green) and the steered 1156\,nm output of the beam clock (orange), and deliberately degrade the stability of the vapor reference for times >0.5\,s by disabling power locks.
In Fig.\,\ref{Fig:Lab} (b) we plot the estimated frequency instability as a modified Allan deviation (ModADEV) produced by carrying out a three cornered hat (TCH) \cite{Gray1974} analysis of the measured frequency comparisons.
For integration times greater than 10\,ms the RB steering signal corrects for frequency instability of the vapor reference, producing clock outputs that integrate down as $2\times 10^{-14}/\sqrt{\tau}$ (black line) to 100\,s, at which point the instability of the vapor reference is suppressed by a factor of 150.  
Due to the instability of the available references, it is difficult to confidently assess the clock performance for longer integration times and this will be explored in future work. 
The Yb beam clock shows improved frequency stability over the Microchip MHM-2020 hydrogen maser over all timescales currently characterized.

The frequency stability data corresponds to the RB spectroscopy shown in Fig.\,\ref{Fig:Spectroscopy} (a).  
For this data we measure a SNR of the fringe of $160$ (in 1\,Hz bandwidth) which gives a predicted clock instability of $\sim 2.1\times 10^{-14}/\sqrt{\tau}$ (Eq. 6), in good agreement with the measured ModADEV.  
We also plot in Fig.\,\ref{Fig:Lab} the expected stability limit due to atom shot noise from atoms contributing to the fringe (dark gray) and after including the additional noise from background atoms (mid gray).
We estimate that we collect 0.5 photons per atom passing through the detection stage for the relevant velocity class.
Including photon shot noise (from both atoms and background scattering) further increases the predicted clock instability (light gray), to a value in reasonable agreement with the measured performance. (For more details see Supplementary Material Sec.\,3)  

\subsection{Field trial} \label{Sec:Sea}

\begin{figure*}[t!] 
\centering
\includegraphics[width=1.\linewidth]{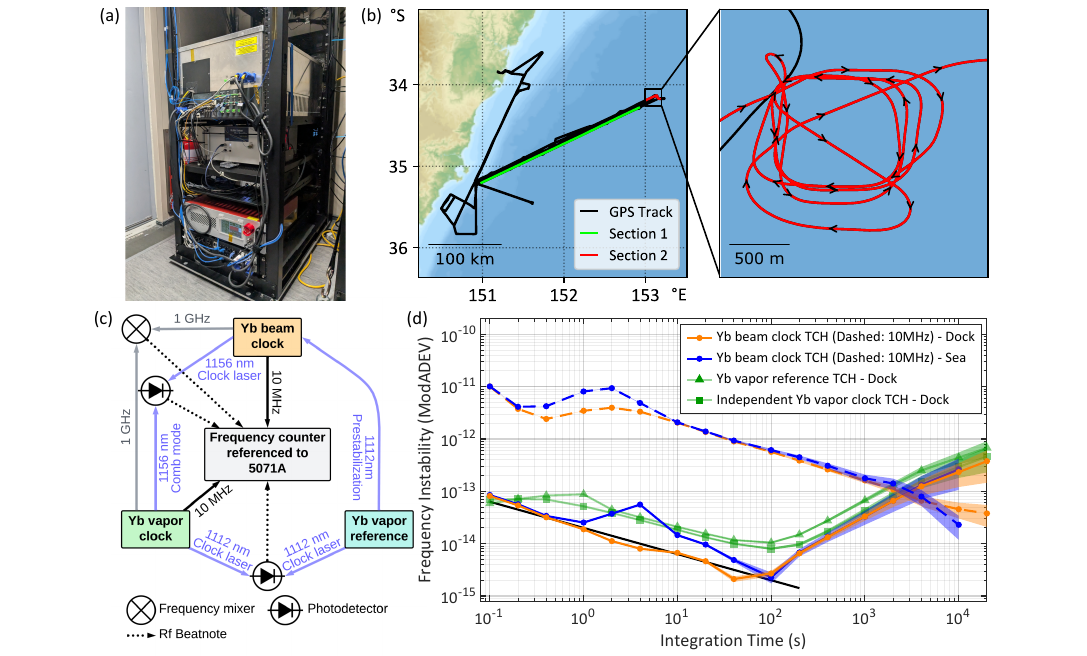}
\caption{\label{Fig:Sea} Field trial of the Yb beam clock. (a) Yb beam clock installed in the ship's air traffic control room, (b) GPS track of 5 days of sailing, with inset showing the zoomed track for the INS calibration maneuver, (c) Schematic of frequency comparisons made during the trial, and (d) Frequency stability (ModADEV) of the Yb beam clock at dock (orange) and at sea (blue) extracted from the TCH measurement, green curves show the extracted stability of the vapor reference (triangles) and Yb vapor clock (squares) at dock.  Measured stability of the Yb beam clock 10\,MHz output referenced to the 5071A (limited by the 5071A stability) at dock (orange, dashed) and at sea (blue, dashed). Black line shows the $2\times 10^{-14}/\sqrt{\tau}$ frequency stability of the Yb beam clock measured in the lab. }
\end{figure*}

The Yb beam clock underwent field testing as part of a defence trial aboard a vessel provided by the Royal Australian Navy in July 2024.  
The trial included initial testing at dock followed by 5 days of testing at sea, and took place off the eastern coast of Australia out of Sydney Harbour (Fig.\,\ref{Fig:Sea} (a) and (b)).  
To take part in the trial, the Yb beam clock was transported from Adelaide to Sydney by commercial freight truck, traveling a total of 1400\,km.  
This required all systems, including the ion pump used to maintain vacuum within the physics package, to be powered down for 10 days. 
On arrival in Sydney the clock and frequency comparison equipment were craned aboard ship and installed into a standard 800\,mm deep rack located within the air traffic control room.
Upon completion of the install, power was returned to the ion pump and the vacuum successfully reached the same base pressure as pre-freight with no need for an additional roughing pump. 

Installation into an existing rack aboard the ship necessitated breaking the system down into its discrete subsystems (as labeled in Fig.\,\ref{Fig:Lab} (a)). 
Due to fixed fiber patch cords within the commercial laser systems, this required the clock and cooling laser heads to be removed from the physics package during transit, and then re-installed once aboard the ship.
Despite this, RB interference fringes were observed without re-alignment of free space optics upon power on, albeit with a degradation in contrast to $40\%$ of the value measured in the lab.
After a small amount of alignment optimization the fringe contrast was restored to the same value as pre-freight. 
This demonstrated the robustness of the physics package as a whole, including the in-vacuum RB interferometer.  

As well as the Yb beam clock, an independent Yb vapor clock \cite{Hilton2024} and Microsemi 5071A cesium clock were installed within the rack as characterization references.  
A schematic of the frequency comparisons made is shown in Fig.\,\ref{Fig:Sea} (c).
In total five separate measurements were logged simultaneously.  
Two of these were difference frequency measurements between the Yb beam clock and Yb vapor clock, between both 1156\,nm and 1\,GHz outputs of these systems. 
An additional difference frequency measurement was made between the 1112\,nm outputs of the Yb vapor clock and the vapor reference within the beam clock. 
Finally the frequency variations of the 10\,MHz outputs from the Yb beam clock and Yb vapor clock were also directly recorded.  
All measurements were made on a zero-dead time K+K FXE frequency counter which was referenced to the Microsemi 5071A. 

\begin{figure}[t] 
\centering
\includegraphics[width=0.6\linewidth]{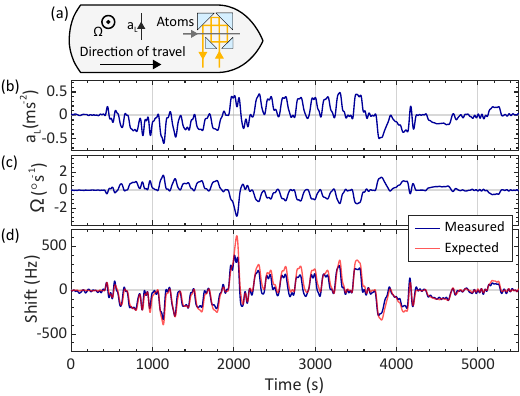}
\caption{\label{Fig:Inertial} Characterization of inertial sensitivity.  (a) Orientation of the RB interferometer and measurement axes for inertial data aboard the ship, (b) Measured linear acceleration, (c) angular velocity, (d) measured and predicted frequency shift of the Yb beam clock (measured against the independent Yb vapor clock at $1156$\,nm).  All plots are filtered to a 0.05\,Hz bandwidth. }
\end{figure}

Initial frequency stability measurements were made before the ship left dock. 
Using the TCH technique the individual performance of the Yb beam clock, Yb vapor clock, and vapor reference were extracted from the modified Allan deviations of the 1\,GHz and 1112\,nm frequency difference measurements (Fig.\,\ref{Fig:Sea} (d)).  
The beam clock (orange curve) shows similar performance to the lab data in Fig.\,\ref{Fig:Lab}, as indicated by the $2\times 10^{-14}/\sqrt{\tau}$ line shown in black.
For times longer than 100\,s a correlated temperature dependence of the Yb vapor clock and vapor reference contaminates the TCH estimation of single-clock performance.  
This is confirmed by the measured frequency stability of the 10\,MHz output of the beam clock (orange, dashed) which, although limited by the microsemi 5071A counter reference, falls below the TCH measurement of the beam clock stability for integration times $>3000$\,s.
The frequency difference measurements made between the optical and 1\,GHz outputs are not limited by the 5071A due to their higher carrier frequencies compared to the 10\,MHz 5071A reference.  

The at-sea phase of the trial provided an important opportunity to characterize the inertial sensitivity of the Yb beam clock.  
The RB spectroscopy technique is inherently sensitive to inertial effects \cite{Borde1989,Riehle1991}. 
Both accelerations and rotations of the system cause the atoms to propagate through different locations of the interrogation beams, producing different optical phases $\phi_i$.
To first order the fractional frequency shift of the clock due to these effects is given by \cite{Borde1989, Jekeli2005}
\begin{equation}
    \begin{aligned}
        &\delta \nu/\nu = S_a a_L + S_r \Omega \\
        &S_a = \frac{L+D}{2 c v_z}, \quad S_r = -\frac{L+D}{c} 
    \end{aligned}
\end{equation}
where $a_L$ is the linear acceleration along the axis of the clock laser, $\Omega$ is the angular velocity about the axis orthogonal to both the atomic beam and clock laser beams, $c$ is the speed of light, $v_z$ is the longitudinal velocity of the atoms, and $L$ and $D$ are the spacing of the clock laser beams as defined in Fig.\,\ref{Fig:Concept}.  
For the parameters of the Yb beam clock, the sensitivities to acceleration and rotation are $S_a=2.45\times10^{-13}/(\mathrm{ms}^{-2})$ and $S_r=8.2\times 10^{-13}/(^{\circ}\mathrm{\text{s}^{-1}})$, respectively. 

These inertial sensitivities lead to degradation of the clock's frequency stability whilst at sea. 
For the segment of data shown in Fig.\,\ref{Fig:Sea} (d) this is observed predominantly for 1--10\,s integration times and can be attributed to wave motion experienced by the ship. 
This data corresponds to the green highlighted portion of the ship's track in Fig.\,\ref{Fig:Sea} (b), during which the ship traveled on a bearing of 234--247$^{\circ}$ with a speed of 9.8--13.4 knots (5.0--6.9\,$\text{ms}^{-1}$). 
As for the at-Dock data, the TCH estimate past 100\,s is unreliable, with the direct 10\,MHz measurement (blue dashed) showing better stability past 3000\,s.
We note that the 10\,MHz measurement remains limited by the performance of the 5071A counter reference whilst at sea, with the increased instability for 0.2--10\,s integration times due to a previously observed acceleration sensitivity of the 5071A itself \cite{Hilton2024}.

The inertial sensitivity is most clearly seen during sections of data where the ship underwent multiple heading changes. 
This is illustrated in Fig.\,\ref{Fig:Inertial} where we show data corresponding to the time in which the ship traveled along the red section of GPS track in Fig.\,\ref{Fig:Sea} (b).  
A clear correlation exists between inertial data recorded on board the ship and the measured frequency difference between the Yb beam clock and Yb vapor clock. 
The linear acceleration was recorded by a sensor (Yoctopuce Yocto-3D) embedded within the beam clock, and we plot the component of the acceleration along the clock laser axis (Fig.\,\ref{Fig:Inertial} (a)).  
The angular velocity of the ship was calculated from the ship's heading as recorded via GPS, and corresponds to rotations around the vertical axis. 
Combining the measured acceleration and angular velocity with the expected inertial sensitivity coefficients $S_a$ and $S_r$ results in an expected frequency shift in good agreement with the measured shift.  

Throughout the trial the beam clock experienced temperature changes of $\pm2.5^{\circ}$C, accelerations of up to 2\,m$\text{s}^{-2}$, and covered over 2000\,km.  
Upon returning to dock, measurements of the fringe contrast showed only a 20$\%$ decrease in contrast over the course of the voyage, during which no realignment of the system was performed.  
Whilst at sea the vapor reference remained continuously locked to the atomic vapor, and all laser frequency offset locks within the beam clock were maintained.  
The lock to the RB spectroscopy within the beam clock unlocked a handful of times due to a lack of range on the power control of the clock laser. 
Through the 7 days of the trial (including 5 days at sea), the clock had a total up-time of $91\%$, an outstanding result for an initial field trial of a complex quantum technology operating within a demanding environment.  
This highlights the inherent compatibility of the beam clock with operation in dynamic settings. 

\section{Discussion} \label{Sec:Discussion}

To our knowledge, the Yb beam clock is the first laser-cooled optical atomic clock to be demonstrated at sea.
This represents a step change for portable optical clock technologies; moving away from vapor-based systems to more controlled, isolated atomic samples inherently reduces environmental sensitivities and allows interrogation of narrower spectroscopy features. 
By interrogating the $^1$S$_0\rightarrow ^3$P$_0$ transition in $^{171}$Yb we are able to produce a 3.5\,kHz-wide clock spectroscopy feature, and additionally benefit from the transition's low sensitivity to magnetic field and black body radiation.  

Our RB fringe period of 7\,kHz was chosen based on the performance of our atom-vapor pre-stabilization reference, however our use of an ultra-narrow transition means there is significant scope to reduce this.  
As portable reference cavities and robust ultra-stable lasers become more developed, narrower fringes could be produced by using additional cooling stages to create a longitudinally slowed beam of atoms for increased Ramsey times. 
As well as improved short-term stability, this would support significant improvements in long-term stability by reducing the frequency sensitivity to interferometer phase $\phi$.

The short-term stability of $2\times 10^{-14}/{\sqrt{\tau}}$ of the beam clock is already competitive with other portable systems \cite{JOKARUS2019, Roslund2024, Perrella2019, Martin2018, Hilton2024}, albeit for significantly higher size-weight-and-power than vapor-based clocks.
A factor of two improvement in the fundamental noise floor would be possible by removing the photon shot noise contribution, either through improvements in collection efficiency or increasing the length of the detection region.
An additional factor of two could be gained by moving to excited state detection to reduce the noise contribution from background atomic flux, bringing the expected shot noise limited performance to $\sim 3\times 10^{-15}/{\sqrt{\tau}}$.
Further optimization may be possible by spectral broadening of the readout laser to tailor the sampled atomic velocity distribution.
With reference to Fig.\,\ref{Fig:Concept} (c), the current velocity-selective readout targets a narrow velocity class with the largest excited flux, however, there is significant excitation of the clock transition spread over a range of velocities. 
Modulating the frequency of the readout laser would allow optimal sampling of this distribution, maximizing the atomic flux contributing to the RB fringe whilst minimizing the background flux. 

The beam clock is considerably more compact than transportable lattice and ion clocks \cite{PTB2018, RIKEN2020, Opticlock2021, Wuhan2020, LENS2014, Bothwell2025}, however these systems demonstrate excellent long-term frequency stability and accuracy.   
The long-term frequency stability of beam clocks is usually limited by path length changes in the interferometer causing variation in the phase $\phi$.
We have mitigated this effect by making use of two displaced retroreflector optics to form the four RB beams (see Fig. \ref{Fig:Chamber}).  This configuration makes $\phi$ insensitive to relative displacements of the interferometer optics thus removing the leading order contribution to phase changes.
Whilst the clock's operation during the field trial confirms the short-term insensitivity of the interferometer, we have not yet characterised the frequency stability of the clock for integration times longer than 400\,s.  
The interferometer also impacts the accuracy of the clock, with $\phi$ determining the zero-crossing of the RB fringe, resulting in frequency shifts of up to the fringe period (7\,kHz).
To date the best fractional frequency stability achieved by a lab-based beam clock is $2\times 10^{-16}$ at 2000\,s \cite{Olson2019}.   
This demonstration made use of two counter-propagating atomic beams to effectively cancel the interferometer phase, as the phase shift is equal but opposite for the two directions.


In addition to this, the phase sensitivity currently limits the clock's frequency stability when the system undergoes significant accelerations and rotations. 
These inertial sensitivities must be reduced for the clock to be suitable for holdover applications on a moving platform.
This can in principal be achieved either via feedforward of measurements from classical inertial sensors \cite{Lautier2014} or by making measurements of the clock system itself in different configurations.
Reversal of the direction of propagation of the atoms leads to an equal but opposite shift due to acceleration, and reversal of the direction of propagation of the clock laser leads to equal but opposite signs for both the acceleration and rotation shifts \cite{Jekeli2005, Salducci2024}.
We have undertaken initial modeling of the RB spectroscopy signal to investigate these techniques based on methods developed in \cite{Strathearn2023}.  
The model additionally quantizes the atomic momentum perpendicular to the lasers, as well as parallel, to allow for imperfect alignment of the laser and atomic beams. 
For counter-propagating atomic beams, aligned to a relatively modest tolerance of $0.01^{\circ}$, we predict a residual fractional frequency shift due to accelerations for the combined error signal to be $<4 \times 10^{-16}/(\text{ms}^{-2})$. 
Alternatively, reversing the direction of propagation of the clock laser, again to a $0.01^{\circ}$ alignment tolerance, would reduce the sensitivity to accelerations to the same value but additionally reduce the sensitivity to rotations to $<1.4 \times 10^{-15}/(^{\circ}\text{s}^{-1})$.
We intend to investigate which combination of atomic beam and k-vector reversal gives optimal clock stability in practice in the presence of both inertial effects and path length changes within the RB interferometer.

\section{Conclusion} \label{Sec:Conclusion}

By combining efficient transverse cooling of a thermal Yb beam with a velocity-selective readout stage and a robust atom-vapor pre-stabilization reference, we are able to make measurements of an ultra-narrow optical transition even within deployed environments.  
Locking to this transition produces a short term-frequency stability for our prototype Yb beam clock that is already competitive with the best available commercial systems \cite{MHM2020, VA_Evergreen, muClock, Tikqer, Tempo} and we expect significant improvements to be possible through optimization of the physics package.
During field trials, the clock operated for multiple days whilst at sea and the measured acceleration and rotation induced frequency shifts showed good agreement with a theoretical model of the system.  
With the implementation of our proposed techniques to reduce the inertial sensitivity, this system has the potential to fill the need for robust, portable high-performance holdover references.  

\section{Back matter}

\begin{backmatter}
\bmsection{Funding}
This research was supported by the Australian Government through the Next Generation Technologies Fund (now managed through ASCA).

\bmsection{Acknowledgment}
The field trail data presented here was collected during a defence trial organized by the Australian Government Department of Defence.
We would particularly like to thank the Defence Science and Technology Group (DSTG) staff who supported the clocks' involvement in the exercise, including Joe Verringer, Mark Baker, and Scott Foster, as well as Joseph Knudsen and the other trial participants. 
We would additionally like to thank Andrew May, Joanne Harrison, 
Ben Sparkes, David Bird, and Anthony Szalbo of DSTG for their support of the University of Adelaide portable atomic clock team.  

We thank the Optofab node of the Australian National Fabrication Facility (ANFF) which utilize Commonwealth and South Australia State Government funding. The authors thank Evan Johnson, Alastair Dowler, and Lijesh Thomas, and the rest of the Optofab team for their technical support.

We also thank Russell Anderson (La Trobe University) for his early contributions to the Yb beam clock project.

\bmsection{Disclosures}
The authors declare that there are no conflicts of interest related to this article.

\bmsection{Data Availability Statement}
Data underlying the results presented in this paper are not publicly available at this time but may be obtained from the authors upon reasonable request.

\end{backmatter}

\bibliography{referencesNew}

\end{document}